\documentclass{CEAS_EuroGNC_2026}

\usepackage[utf8]{inputenc}

\usepackage{graphicx}
\usepackage{amsmath}
\usepackage[version=4]{mhchem}
\usepackage{siunitx}
\usepackage{array}
\usepackage{longtable}%tabulary
\usepackage{framed}
\usepackage[most]{tcolorbox}
\newtcolorbox{tcbdoublebox}[1][]{%
  enhanced jigsaw,
  sharp corners,
  colback=white,
  borderline={1pt}{-2pt}{black},
  fontupper={\setlength{\parindent}{20pt}},
  #1
}

\usepackage{color}
\usepackage{epstopdf} 
\usepackage{amsfonts}
\usepackage{xcolor}
\usepackage{algorithm}
\usepackage{epsfig}
\usepackage{tabularx}
\usepackage{booktabs} % for \specialrule command
\usepackage{algorithm, algorithmicx, algpseudocode}
\usepackage{multirow}
\newcommand{\nx}{n_\mathrm{x}}
\newcommand{\dnu}{n_\mathrm{u}}

\newcommand{\D}{\mathcal{D}}
\newcommand{\Real}{\mathbb{R}}

\title{Data-driven Learning of LPV Surrogate Models \\ of Fuel Sloshing}
\papernumber{103}% INSERT THE PAPER NUMBER THAT THE ORGANISATION TEAM PROVIDES YOU

\ifpdf
\hypersetup{pdfinfo={
Title={Data-driven Learning of LPV Surrogate Models of Fuel Sloshing},% replace with your paper title (without special maths symbols, for instance H-infinity instead of $H_\infty$
Author={E. Javier Olucha, Valentin Preda, Amritam Das, Roland Tóth},%insert all authors with the following format (here for 3 authors) {First Author, Second Author, and Third Author}
Keywords={Surrogate modelling, Fuel sloshing, Smoothed-Particle Hydrodynamics, Linear-Parameter Varying} % insert keywords and separate them with commas
}}
\fi

\begin{document}
\thispagestyle{fancy}% do not change

\maketitle% do not change
\pagestyle{empty} % do not change

% Argument is the width that is reserved for the author names. Authors MAY slightly customise this value to improve the layout (e.g. to avoid unaestetic breaking of their names)
\begin{authorList}{4cm} % <- CUSTOMISE WITH IF NEEDED (default 4cm)
\addAuthor{E. Javier Olucha \orcidlink{0009-0001-6731-6523}}{PhD candidate, Control Systems Group, Eindhoven University of Technology~\rorlink{https://ror.org/02c2kyt77}, Eindhoven, The Netherlands. \emailAddress{e.j.olucha.delgado@tue.nl}}
\addAuthor{Valentin Preda \orcidlink{0000-0003-1394-9832}}{GNC System Engineer, Department of Guidance, Navigation and Control, European Space Agency~\rorlink{https://ror.org/03h3jqn23}, Noordwijk, The Netherlands. \emailAddress{valentin.preda@esa.int}}
\addAuthor{Amritam Das \orcidlink{0000-0001-8494-8509}}{Assistant Professor, Control Systems Group, Eindhoven University of Technology~\rorlink{https://ror.org/02c2kyt77}, Eindhoven, The Netherlands. \emailAddress{am.das@tue.nl}}
\addAuthor{Roland Tóth \orcidlink{0000-0001-7570-6129}}{Full Professor, Control Systems Group, Eindhoven University of Technology~\rorlink{https://ror.org/02c2kyt77}, Eindhoven, The Netherlands. \emailAddress{r.toth@tue.nl}\newline
Senior Research Fellow, Systems and Control Laboratory, HUN-REN SZTAKI~\rorlink{https://ror.org/04w6pnc49}, Budapest, Hungary. \emailAddress{toth.roland@sztaki.hun-ren.hu}}
\end{authorList}
\justifying

\begin{abstract}
This paper aims to enhance the efficiency of validation and verification campaigns involving fuel sloshing phenomena.
Our first contribution is the development of an open-source, high-fidelity and computationally efficient two-dimensional smoothed-particle hydrodynamics-based fuel sloshing simulator that reproduces the dynamics of a spacecraft with a partially filled tank with liquid propellant. Implemented in \textsc{Python} using \textsc{Jax}, the simulator leverages GPU parallelization and supports automatic differentiation, enabling rapid generation of simulation data and system linearizations for general surrogate modelling purposes. 
Our second contribution is the demonstration of a practical methodology for constructing surrogate models of fuel sloshing from input--output data generated by the simulator, targeting rapid simulation and model-based control applications. The surrogate model employs a \emph{linear parameter-varying} (LPV) state-space structure with affine dependence on the scheduling variables, providing an accurate yet computationally efficient approximation of the sloshing dynamics.
The capabilities of the proposed approach are demonstrated through closed-loop simulations of a rigid spacecraft with a partially filled fuel tank for two manoeuvre profiles under zero-gravity conditions. The identified surrogate enables simulations that are two orders of magnitude faster than the high-fidelity model.
\end{abstract}
% CUSTOMIZE KEYWORDS
% do not use CEAS or EuroGNC as keyword and make sure that these keywords match the
% ones provided in the "pdfinfo" provided about 30 lines earlier 
\section*{Nomenclature}
{\renewcommand\arraystretch{1.0}
\noindent\begin{longtable*}{@{}l @{\quad=\quad} l@{}}
GPU & Graphics processing unit \\
% CPU & Central processing unit \\
SPH & Smoothed-Particle Hydrodynamics \\
SS & State-Space \\
LTI & Linear-Time Invariant \\
LPV & Linear-Parameter Varying \\
CoM & Centre of mass \\
$\mathcal{B}$ & Rigid body representing the spacecraft \\
$m, J$ & Mass and inertia of the spacecraft rigid body, respectively \\
$\mathcal{R}_{\mathcal{W}}, \mathcal{R}_{\mathcal{B}}$ & Inertial world and body-fixed reference frames, respectively\\
$r, \theta$ & Linear and angular position of the spacecraft at the CoM with respect to $\mathcal{R}_{\mathcal{W}}$ \\
% $u, y$ & Input and output vectors, respectively \\
% $k$ & Discrete time \\
$\mathcal{D}_N$ & Data set \\
$\vartheta$ & Surrogate model parameters \\
$m_i, \rho_i, A_i$ & Mass, density and field quantity of particle $j$, respectively \\
$W$ & Radially symmetric kernel function with compact support \\
$h$ & Smoothing length of a kernel function $W$ \\
$\rho_0, \alpha$ & Rest density and viscous factor of the fluid, respectively \\
$\beta, \gamma_1$ & Fluid-boundary viscous factor and boundary correcting factor, respectively \\
\end{longtable*}}
% % % % % % % % % % % % % % % % % % % % % % % % % % % % % % % % % % % % % % % % 
% MAIN MATTER
\section{Introduction~\label{sec:intro}}
In aerospace systems such as satellites, launchers, and upper stages, liquid propellants are used for vehicle propulsion when executing various manoeuvres such as flight-to-orbit, rendezvous, or station keeping. Due to propellant consumption, these systems typically operate with partially filled tanks. Consequently, during manoeuvres, the propellant experiences motion relative to the tank walls, a phenomenon known as \emph{fuel sloshing}~\cite{abramson1966, neer1972}. The motion of the fluid and its interaction with the fuel tank boundaries induce a dynamic coupling between the fluid and the rigid body, effecting the overall spacecraft dynamics. These fluid--solid interactions can degrade pointing accuracy and may even lead to instability if not properly accounted for~\cite{ibrahim2005, liu2018, simonini2024}. Moreover, as spacecrafts become more agile and lightweight, the impact of fuel sloshing becomes increasingly critical and must be carefully addressed during both controller design and \emph{validation and verification} (V\&V) campaigns.

Accurate representation of fuel sloshing requires \emph{high-fidelity} models capable of capturing complex physical effects such as free-surface deformation, viscous damping and the interaction between the fluid and the solid boundaries. However, such models are computationally expensive, high-dimensional, and strongly nonlinear, and they are rarely available in an analytic \emph{state-space} (SS) representation. As a result, they cannot be directly integrated into conventional control design or system analysis toolchains, and their use is typically restricted to offline analyses. Furthermore, their computational complexity limits the efficiency and cost-effectiveness of V\&V campaigns and hinders rapid system-level evaluations during early design stages.

To circumvent these challenges, \emph{surrogate models} can be employed to approximate the effects of fuel sloshing. A surrogate model is a simplified representation of a high-fidelity model with minimal complexity, but with sufficient system information and adequate model structure to reach a specific utilization objective. The development of suitable fuel sloshing surrogate models is therefore a key enabler for efficiently integrating accurate fluid--solid interactions into early-stage engineering processes.

In the context of fuel sloshing, two main classes of models are primarily used in the literature: (i) high-fidelity \emph{computational fluid dynamics} (CFD)~\cite{vreeburg2005, lazzarin2014, dalmon2018, dalmon2019} or \emph{smoothed particle hydrodynamics} (SPH)~\cite{hahn2018, atif2019, kotsarinis2023} representations that are accurate but suffer from the drawbacks aforementioned, and (ii) low-fidelity, control-oriented analytical or empirical models, often based on mechanical analogies~\cite{enright1994, jang2013, gasbarri2016, oluchadelgado2023, rodrigues2023} with low complexity but may fail to capture the sloshing phenomena with sufficient accuracy. An intermediate class of surrogate models that balances physical fidelity and computational tractability remains largely missing for early-stage engineering tasks and downstream V\&V campaigns.

The systematic extraction of surrogate models typically relies on either analytical descriptions of the underlying system dynamics or on datasets composed of system inputs, measured outputs, state trajectories, or linearized dynamics around selected operating points. For systems with fuel sloshing, analytical descriptions are rarely available, whereas data can be generated using suitable physics-based simulators. However, despite the availability of several fuel sloshing simulators, open-source tools capable of efficiently generating informative datasets in terms of system trajectories or linearizations are mostly unavailable.
Such datasets are essential for constructing surrogate models that approximate the input--output behaviour, and potentially the internal dynamics, of high-fidelity simulations. Therefore, developing an efficient fuel sloshing simulator capable of generating these datasets and establishing a consistent methodology for surrogate model extraction are crucial steps towards improving the overall efficiency of aerospace engineering processes.

In this paper, two main contributions are presented. The first is the development of an ultra-fast, open-source, high-fidelity \emph{two-dimensional} (2D) SPH-based fuel sloshing simulator that: (i) leverages \emph{graphics processing units} (GPU) for parallel computation of particle interactions; (ii) exploits \emph{just-in-time} (JIT) compilation provided by \textsc{Jax}~\cite{bradbury2018}; and (iii) supports \emph{automatic differentiation}~\cite{maclaurin2015} to efficiently obtain linearized dynamics around arbitrary operating points. The simulator provides a flexible framework for exciting a spacecraft containing a tank partially filled with fluid propellant and enables systematic dataset generation for surrogate modelling purposes. 
The second contribution is the demonstration of a practical methodology for learning \emph{linear parameter-varying} (LPV) surrogate models of fuel sloshing dynamics using data generated by the simulator. In particular, we show how input--output data can be used to identify a self-scheduled LPV state-space model with affine dependence on the scheduling. The resulting surrogate model provides an accurate yet computationally efficient approximation, making it suitable for rapid simulation and integration into control design and V\&V workflows.

This paper is structured as follows. First, we introduce the surrogate modelling problem for the fuel sloshing scenario in Section~\ref{sec:problem_def}, and we detail the development of the high-fidelity SPH-based fuel sloshing simulator in Section~\ref{sec:simulator}. Section~\ref{sec:surrogate} presents the LPV surrogate modelling approach, and we demonstrate the capabilities of the SPH-based fuel sloshing simulator together with the performance of the identified LPV surrogate in Section~\ref{sec:results}. Finally, conclusions are drawn in Section~\ref{sec:conclusions}.
\section{Problem definition~\label{sec:problem_def}}
\subsection{Definition of the fuel sloshing scenario\label{sec:scenario}}
\begin{figure}[t]
    \centering
    \includegraphics[width=0.7\linewidth]{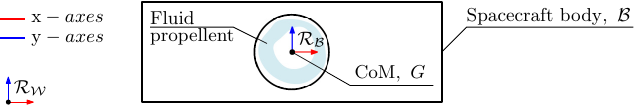}
    \caption{Sketch of the fuel sloshing scenario considered in this paper, where $\mathcal{R}_\mathcal{W}$ denotes the inertial world reference frame, and $\mathcal{R}_\mathcal{B}$ denotes a local body frame attached at the CoM $G$ of the spacecraft rigid body $\mathcal{B}$.}
    \label{fig:fuel_sloshing_sketch}
\end{figure}

In this paper, we consider a two-dimensional spacecraft containing a partially filled fuel tank, as illustrated in Fig.~\ref{fig:fuel_sloshing_sketch}. The spacecraft is modelled as a rigid body, denoted by $\mathcal{B}$, with \emph{centre of mass} (CoM) located at point $G$. The fuel tank is rigidly attached to the spacecraft and contains fluid propellant that can move freely within the tank boundaries.
A body-fixed reference frame attached to the spacecraft CoM is defined as $\mathcal{R}_{\mathcal{B}} = (G; x_G, y_G)$, while an inertial reference frame is denoted by $\mathcal{R}_{\mathcal{W}}$. Unless otherwise specified, all positions, velocities, and accelerations are expressed in the inertial frame $\mathcal{R}_{\mathcal{W}}$.

The translational motion of the spacecraft at a point $Q$ is described by the position vector
$r(t) = [ r_{\mathrm x}(t) \ r_{\mathrm y}(t)]^\top \in \mathbb{R}^2$,
and its orientation by the rotation angle $\theta(t) \in \mathbb{R}$. Their time derivatives $\dot{r}(t)$, $\ddot{r}(t)$ and $\dot{\theta}(t)$, $\ddot{\theta}(t)$ denote the corresponding translational and rotational velocities and accelerations.
The system is subjected to input forces and torque applied at a point $P$, representing the action of spacecraft thrusters, and the input vector is defined as $u(t) = [u_{\mathrm x} \ u_{\mathrm y} \ \tau]^\top(t) \in \mathbb{R}^3$.
For simplicity, we assume that $P$ and $Q$ are located at the CoM of the spacecraft, i.e., $P=G$ and $Q=G$.
The measured output of the system, expressed with respect to the spacecraft CoM, is then defined as $y(t) = [r_{\mathrm x} \ r_{\mathrm y} \ \theta \ \dot{r}_{\mathrm x} \ \dot{r}_{\mathrm y} \ \dot{\theta}]^\top \in \mathbb{R}^6$.
The sloshing motion of the fluid induces hydrodynamic forces and torques on the spacecraft, thereby coupling the rigid-body dynamics with the internal fluid motion. The mass and temperature of both the fluid and the satellite are assumed constant.
\subsection{Problem formulation and objectives\label{sec:objectives}}
Based on the fuel sloshing scenario described in Section~\ref{sec:scenario}, the two complementary challenges identified in Section~\ref{sec:intro} are formalized as the following objectives.
\begin{enumerate}
    \item \textit{\textbf{High-fidelity fuel sloshing simulator.}} 
    The first objective is the development of a computational model $\Sigma$ capable of reproducing the coupled spacecraft-fuel sloshing dynamics with high physical fidelity, of the form:
    \begin{equation} \label{eq:NLmodel}
        \Sigma: \left\{ \ \ 
        \begin{aligned}
            \dot{x}(t) &= f(x(t), u(t)) , \\
            y(t) &= h(x(t), u(t)),
        \end{aligned}\right.
    \end{equation}
where $t \in \Real$ is time, $x(t) \in \mathbb{X} \subseteq \Real^{\nx}$ with $\nx \in \mathbb{Z}$ is the state variable, $u(t) \in \mathbb{U} \subseteq \Real^{3}$ the input of the system, $y(t) \in \mathbb{Y} \subseteq \Real^6$ is the measured output, corresponding to the translational and rotational motion of the system as defined in Section~\ref{sec:scenario}. The functions $f:\Real^{\nx} \times  \Real^{3} \to \Real^{\nx}$ and $h:\Real^{\nx} \times  \Real^{3} \to \Real^6$ represent the (nonlinear) evolution of the coupled rigid-body and fluid dynamics. 
For any initial condition $x(t_0) = x_0$ and input trajectory $u(t)$, the solutions of~\eqref{eq:NLmodel} are forward complete and unique for all $t\geq t_0$.
The model $\Sigma$ serves as an efficient data-generation platform for the development and validation of surrogate modelling methods. Therefore, it must be able to rapidly generate data in terms of input, output and state trajectories of the satellite and/or the fluid, as well as local linearizations at arbitrary operating conditions.
    
    \item \textit{\textbf{Simulation-oriented fuel sloshing surrogate model.}} 
    For the second objective,
    we assume that we have collected a data set $\D_N$ from the high-fidelity computational model~\eqref{eq:NLmodel}:
    \begin{equation}\label{eq:dataset}
    \D_N = \{\left(u_k, y_k\right)\}_{k=0}^{N-1},
    \end{equation}    
    where $\{y_k\}_{k=0}^{N-1}$ is the response sampled at time instances $t = k T_{\mathrm s}$, i.e., $y_k = y(k T_{\mathrm s})$, with $k \in \mathbb{Z}^ +$ and $T_{\mathrm s}$ the sampling time, for a given excitation sequence $\{u_k\}_{k=0}^{N-1}$. We also assume the fluid is initialized in stationary conditions. 
    
    Then, based on $\mathcal{D}_N$, the goal is to identify a discrete-time surrogate model $S$ that captures the fuel sloshing system dynamics, of the form:
    \begin{equation} \label{eq:surrogate}
    S: \left\{ 
    \ \ \begin{aligned}
        \hat{x}_{k+1} &= \hat{f}(\hat{x}_k, u_k, \vartheta), \\
        \hat{y}_k &= \hat{h}(\hat{x}_k, u_k, \vartheta),  
    \end{aligned}\right.
    \end{equation}
    where $k \in \mathbb{Z}^+$ denotes the discrete time-step, $\hat{x}_k \in \Real^{\hat{n}_{\mathrm x}}$ with $\hat{n}_{\mathrm x} \ge 1$ is the surrogate state, $\vartheta$ the surrogate model parameters and $\hat{f}:\Real^{\hat{n}_{\mathrm x}} \times \Real^{\dnu} \times \Real^{n_{\vartheta}} \to \Real^{\hat{n}_{\mathrm x}}$ and $\hat{h}:\Real^{\hat{n}_{\mathrm x}} \times \Real^{\dnu} \times \Real^{n_{\vartheta}} \to \Real^{{n}_{\mathrm y}}$ are functions parametrized by $\vartheta$. The surrogate model $S$ aims at: (i) minimizing the criterion
    \begin{equation}\label{eq:pem}
        J_{\D_N}(\vartheta, \hat{x}_0) = \frac{1}{N}\sum_{k=0}^{N-1} \|y_k - \hat{y}_k\|_2^2,
    \end{equation}
    with $\hat{x}_0 \in \Real^{\hat{n}_{\mathrm x}}$ as the initial condition, and (ii) the parametrization of $\hat{f}$ and $\hat{h}$ ensures a computationally efficient model evaluation.
\end{enumerate}
\section{SPH-based high-fidelity simulator for fuel sloshing with GPU acceleration and automatic differentiation~\label{sec:simulator}}
This section addresses the first objective in Subsection~\ref{sec:objectives}, namely the development of a high-fidelity and computationally efficient model of the fuel sloshing dynamics.
\subsection{Modelling fluid dynamics with the SPH framework\label{sec:fluiddynamics}}
Smoothed-particle hydrodynamics, initially developed by Gingold and Monaghan~\cite{10.1093/mnras/181.3.375} and Lucy~\cite{lucy1997} for the simulation of astrophysical problems, is a computational method for simulating the mechanics of continuum media. By construction, it is a mesh free Lagrangian method and therefore it is suited to approach problems like free surface flows. The lack of mesh simplifies the parallelization of computations. These features make the SPH approach ideal for developing efficient simulators of fuel sloshing.

The SPH method approximates continuous media with a set of discrete moving elements, referred to as particles. These particles interact through a radially symmetric kernel function $W$ with compact support radius $H$ and ``smoothing length'' $h$. Then, a physical quantity can be approximated at any position $r$ based on the relevant quantities of the other particles. For this, the interpolation of a field quantity $A$ is based on
% \TR{[What is the domain of the integral below]}
\begin{equation}\label{eq:integralinterp}
    A(r) = \int_{\Omega} A(r') W(\|r - r'\|; h) \mathrm{d}r',
\end{equation}
where $\mathrm d r'$ denotes a differential volume element~\cite{monaghan_2005}, 
and in two dimensions the integration domain is the compact support $\Omega = \{\,r' \in \mathbb{R}^2 \ | \ \exists \, r \in \Real^2 \ \text{s.t.} \ \|r' - r\| \le H\,\}$.
If the kernel is normalized, i.e.,
\begin{equation}\label{eq:normkernel}
    \int_{\Omega} W(r)\mathrm{d}r =1,
\end{equation}
the interpolation~\eqref{eq:integralinterp} is of second order accuracy~\cite{muller2003}. The integral in Eq.~\eqref{eq:integralinterp} is then approximated using a Riemann summation over the particles:
\begin{equation}\label{eq:suminterp}
    A(r) = \sum_j \frac{m_j}{\rho_j} A_j W(\|r_{i,j}\|; h),
\end{equation}
where $m_j$, $\rho_j$, and $A_j$ denote the mass, the density and the field quantity  for a particle $j$, $r_{i,j} = r_i - r_j $, and the summation over $j$ includes all particles\footnote{In practice, it is sufficient to include the neighbouring particles that lie within the support of the kernel.}. Based on this, the density $\rho_i$ of particle $i$ can be obtained as:
\begin{equation}\label{eq:density}
    \rho_i = \sum_j m_j W_{i,j},
\end{equation}
where $W_{i,j}$ is a shorthand notation for $W(\|r_i - r_j\|; h)$.
In general, for incompressible isothermal viscous fluids, the SPH acceleration equation is given by
\begin{equation}
   m_i \ddot{r}_i = -F_i^{\mathrm{pressure}} + F_i^{\mathrm{viscous}} + F_i^{\mathrm{external}},
\end{equation}
where $\ddot{r}_i$ is the acceleration of particle $i$, and the forces $F_i^{\mathrm{pressure}}$, $F_i^{\mathrm{viscous}}$ and $F_i^{\mathrm{external}}$ are related to the conservation of momentum, the fluid viscosity and the external forces, respectively. There exist multiple approaches to model these forces in the literature. We choose the following approach for $F_i^{\mathrm{pressure}}$, as, for a fixed $h$, the linear and angular momentum are exactly conserved~\cite{monaghan_2005}:
\begin{equation}\label{eq:momentum}
    F_i^{\mathrm{pressure}} = m_i \sum_j m_j \left(\frac{P_i}{\rho_i^2} + \frac{P_j}{\rho_j^2}\right) \nabla_i W_{i,j},
\end{equation}
where $P_i$ is the pressure of particle $i$ that can be computed with the so-called \emph{state equation}~\cite{koschier_2019}:
\begin{equation}\label{eq:EOS}
    P_i = k(\rho_i - \rho_0),
\end{equation}
where $k$ and $\rho_0$ represent stiffness and rest density constants, respectively, and $\nabla_i W_{i,j}$ denotes the gradient taken with respect to the coordinates of particle $i$:
\begin{equation*}
    \nabla_i W_{i,j} = \frac{\partial W_{i,j}}{\partial \|r_{i,j}\|}\frac{r_{i,j}}{\|r_{i,j}\|}, \qquad \nabla_i W_{i,j} = -\nabla_j W_{i,j}.
\end{equation*}
Note that fluids whose pressure can be accurately described by a state equation such as~\eqref{eq:EOS} are commonly referred to as \emph{weakly compressible}.
Alternative formulations that enforce incompressibility exist~\cite{solenthaler2009,bodin2012,macklin2013,ihmsen2014}, but these involve computationally demanding iterative procedures.

Lastly, the inclusion of viscous forces involves the discretization of the Laplacian differential operator, but this discretization leads to a poor estimate~\cite{koschier_2019}. Therefore, we make use of the ``artificial viscosity'' approach introduced in~\cite{monaghan1983}:
\begin{equation}\label{eq:viscous}
    F_i^{\mathrm{viscous}} = m_i \sum_j \, m_j \frac{2 \,\alpha \, h}{\rho_i + \rho_j} \frac{\dot{r}_{i,j} \cdot r_{i,j}}{\|r_{i,j}\|^2 + \epsilon \, h^2} \nabla_i W_{i,j}
\end{equation}
where $\alpha$ represents the viscous factor of the fluid, and $\epsilon \sim 0.01$ is introduced to prevent a numerical singularity when $r_{i,j}=0$.

\subsection{Modelling the boundaries of the fuel tank with ghost particles\label{sec:ghostparticles}}
We model the boundaries of the fuel tank using a particle based strategy, where the geometry of the boundary is populated by a single layer of uniformly distributed \emph{ghost} particles. These ghost particles are incorporated into the computation of field properties of fluid particles, and will serve to define the forces and effects resulting from the interaction between the fluid and the tank walls, as well as to prevent \emph{tunnelling}, i.e., fluid particles  penetrating the boundary. 

\begin{figure}[t]
    \centering
    \includegraphics[width=0.5\linewidth]{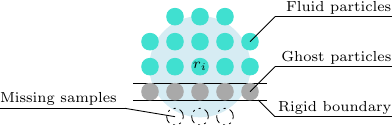}
    \caption{Modelling of the tank boundary with a single layer of uniformly distributed ghost particles. For fluid particles near the boundary, the computation of field quantities may omit contributions from missing samples, i.e., neighbouring particles lying outside the simulated domain but within the kernel support.}
    \label{fig:boundary_sketch}
\end{figure}

Each ghost particle is then considered to be a static fluid particle with respect to the fuel tank frame $\mathcal{R}_{\mathcal{B}}$, i.e., their relative position w.r.t. $\mathcal{R}_{\mathcal{B}}$ does not change over time, replicating the geometry of the boundary. In addition, ghost particles inherit the properties of fluid particles, and the volume of ghost particles must be equal or lower than the fluid counterpart to prevent tunnelling. When including ghost particles, the density computation in Eq.~\eqref{eq:density} is modified to
\begin{equation}
    \rho_i = m_i \left(\sum_{i_f}W_{i, i_f} + \sum_{i_g}W_{i, i_g} + \sum_{i_m}W_{i, i_m} \right),
\end{equation}
where $i_f$, $i_g$ and $i_m$ are the indices that refer to fluid, ghost and missing particles, respectively. In particular, missing particles are the ones that lie within the kernel support, but are not included in the single layer realization of the boundary, as illustrated in Figure~\ref{fig:boundary_sketch}. Therefore, fluid particles close to a boundary miss the contribution from missing particles, consequently underestimating the field quantities.
A practical approach to tackle this issue is to encode the contribution from missing ghost particles in a correcting factor $\gamma_1$:
\begin{equation}\label{eq:density_update}
    \rho_i = m_i \left(\sum_{i_f}W_{i, i_f} + \gamma_1\sum_{i_g}W_{i, i_g} \right),
\end{equation}
where $\gamma_1$ can be analytically estimated~\cite{koschier_2019} as
\begin{equation}\label{eq:gamma1}
    \gamma_1 = \frac{\frac{\rho_i}{m_i} - \sum_{i_f} W_{i, i_f}}{\sum_{i_f} W_{i, i_b}}.
\end{equation}
Then, the pressure force that a ghost particle $j_g$ exerts onto a fluid particle $i_f$ can be expressed as
\begin{equation}\label{eq:pressure_b2f}
    F_{i_f \leftarrow j_g}^{\mathrm{pressure, g2f}} = 2 m_{i_f} m_{j_g} \frac{P_{i_f}}{\rho_{i_f}^2}\nabla_{i_f} W_{i_f, j_g},
\end{equation}
whereas ghost particles inherit the physical properties of the fluid, $m_{i_f} = m_{j_g}$, and the symmetric pressure force from a fluid particle $i_f$ to a ghost particle $j_g$ is
\begin{equation}\label{eq:pressure_f2b}
    F_{i_f \leftarrow j_g}^{\mathrm{pressure, g2f}} = -F_{j_g \leftarrow i_f}^{\mathrm{pressure, f2g}}.
\end{equation}

Moreover, as proposed in~\cite{akinci_2012}, the viscous force between the a fluid particle $i_f$ to a ghost particle $j_g$ can be approximated by
\begin{equation}\label{eq:viscous_b2f}
    F_{i_f \leftarrow j_g}^{\mathrm{viscous, g2f}} = m_{i_f} m_{j_g} \frac{2 \beta}{\rho_{i_f} + \rho_{j_g}} \frac{\min{(v_{i_f j_g} \cdot r_{i_f j_g}, 0)}}{\|r_{i_f j_g}\|^2 + \epsilon \, h^2} \nabla_{i_f} W_{i_f, j_g},
\end{equation}
where $\beta$ is a coefficient that represents the viscosity between the fluid and the solid boundary, $\rho_{i_f} = \rho_{j_g}$ for the reasons stated above, and the symmetric viscous force from a fluid particle $i_f$ to a ghost particle $j_g$ is
\begin{equation}\label{eq:viscous_f2b}
    F_{i_f \leftarrow j_g}^{\mathrm{viscous, g2f}} = -F_{j_g \leftarrow i_f}^{\mathrm{viscous, f2g}}.
\end{equation}

\subsection{Rigid body dynamics and kinematic coupling between the ghost particles and the spacecraft\label{sec:tankdynamics}}
To satisfy our first objective, the fuel sloshing simulator must capture the rigid-body dynamics of the spacecraft, including the forces exerted by the fluid on it. 
The Newton--Euler equations of motion in two-dimensional space~\cite{ardema2005}, expressed with respect to $\mathcal{R}_{\mathcal{W}}$, are given by
\begin{equation}\label{eq:tankdynamics}
\begin{aligned}
    m \, \ddot{r} &= \sum_{j_g} \sum_{i_f} \left(F^{\mathrm{pressure,f2g}}_{{j_g}\leftarrow {i_f}} + F^{\mathrm{viscous,f2g}}_{{j_g}\leftarrow {i_f}} \right) + F^{\mathrm{external}}, \\
    J \, \ddot{\theta} &= \sum_{j_g}  (r_{j_g} - r) \times \left(\sum_{i_f} F^{\mathrm{pressure,f2g}}_{{j_g}\leftarrow {i_f}} + \sum_{i_f} F^{\mathrm{viscous,f2g}}_{{j_g}\leftarrow {i_f}} \right) + T^{\mathrm{external}},
\end{aligned}  
\end{equation}
where $m$ and $J$ are the respective mass and inertia of the spacecraft rigid body, and $F^{\mathrm{external}}$ and $T^{\mathrm{external}}$ are the external force and torque applied at the CoM. The interaction forces $F^{\mathrm{pressure,f2g}}_{{j_g}\leftarrow {i_f}}$ and $F^{\mathrm{viscous,f2g}}_{{j_g}\leftarrow {i_f}}$ are given by Eqs.~\eqref{eq:pressure_f2b} and~\eqref{eq:viscous_f2b}, respectively.

While the forces acting on ghost particles due to the fluid are accounted for in~\eqref{eq:tankdynamics}, ghost particles are not automatically constrained to follow the rigid-body motion of the tank. Therefore, an explicit kinematic coupling is introduced. As described in Section~\ref{sec:ghostparticles}, the position of ghost particles remain fixed in the body frame $\mathcal{R}_{\mathcal{B}}$. Let $r_{j_g}^{\mathcal{R}_{\mathcal{B}}}$ denote the position of the $j$-th ghost particle expressed in $\mathcal{R}_{\mathcal{B}}$. Then, the kinematic update of a ghost particle is given by
\begin{equation}\label{eq:kinematicghost}
\begin{aligned}
    r_{j_g} &= R(\theta) r_{j_g}^{\mathcal{R}_{\mathcal{B}}} + r, \\
    \dot{r}_{j_g} &= \dot{r} + \dot{\theta} \times (r_{j_g} - r),
\end{aligned} \qquad \text{where} \qquad R(\theta) = \begin{bmatrix}
        \cos \theta & -\sin \theta \\ \sin \theta & \cos \theta 
    \end{bmatrix}.
\end{equation}

\subsection{Fuel sloshing simulation engine}
Building upon the formulation introduced in Subsections~\ref{sec:fluiddynamics},~\ref{sec:ghostparticles} and~\ref{sec:tankdynamics}, we now combine these ingredients to assemble the state transition function $\dot{x}(t) = f(x(t), u(t))$ of the high-fidelity computational fuel sloshing model~\eqref{eq:NLmodel}. The state vector is defined as
\begin{equation*}
    x(t) = \begin{bmatrix}
        {r}(t) &  {\theta}(t)  & r_{i_f}(t)  & \dot{r}(t)  & \dot{\theta}(t)  & \dot{r}_{i_f}(t)
    \end{bmatrix}^\top,
\end{equation*}
and the corresponding state transition function is detailed in Algorithm~\ref{alg:state_evol_fun}. The output mapping $y(t) = h(x(t), u(t))$ of the high-fidelity model~\eqref{eq:NLmodel} is obtained by retrieving the desired quantities of interest (e.g., pose of the tank, fluid-boundary forces or fluid density) directly from Algorithm~\ref{alg:state_evol_fun}. Lastly, in the SPH literature, symplectic integration schemes are preferred over non-symplectic alternatives for the time integration step of particle states, as they conserve angular momentum and ensure reversibility~\cite{monaghan_2005}. Accordingly, we employ a first-order symplectic Euler integration scheme~\cite[Chapter~6]{hairer2006}.
\begin{algorithm}[t]
    \caption{State transition function $\dot{x}(t) = f(x(t), u(t))$ of the high-fidelity computational fuel sloshing model~\eqref{eq:NLmodel}}
    \label{alg:state_evol_fun}
    \textbf{Inputs:} The state $x(t) = [{r} \ \ {\theta} \ \ {r}_{i_f} \ \ \dot{r} \ \ \dot{\theta} \ \ \dot{r}_{i_f}]^\top(t)$ at any time instance $t$.\\
    \textbf{Output:} $\dot x(t) = [\dot{r} \ \ \dot{\theta} \ \ \dot{r}_{i_f} \ \ \ddot{r} \ \ \ddot{\theta} \ \ \ddot{r}_{i_f}]^\top(t)$.\\
    \begin{algorithmic}[1]
        \For{each ghost particle $j_g$}
        \State update the position $r_{j_g, k}$ and velocity $\dot{r}_{j_g, k}$ w.r.t. $\mathcal{R}_{\mathcal{W}}$ with Eq.~\eqref{eq:kinematicghost}.
        \EndFor
        \For{each fluid particle ${i_f}$}
            \State compute the density $\rho_{i_f}(t)$  with Eq.~\eqref{eq:density_update}.
            \State compute the pressure $P_{i_f}(t)$ with Eq.~\eqref{eq:EOS}.
            \State compute $F_{i_f}^{\mathrm{pressure}}(t)$ and $F_{i_f}^{\mathrm{viscous}}(t)$ with Eqs.~\eqref{eq:momentum} and~\eqref{eq:viscous}, respectively.
            \State compute \mbox{$F_{i_f}^{\mathrm{g2f}}(t) = \sum_{j_g} (F_{{i_f} \leftarrow j_g}^{\mathrm{pressure, g2f}}(t) +  F_{{i_f} \leftarrow j_g}^{\mathrm{viscous, g2f}}(t))$} using Eqs.~\eqref{eq:pressure_b2f} and~\eqref{eq:viscous_b2f}.
            \State compute $\ddot{r}_{i_f}(t) = m_{i_f}^{-1}(-F_{i_f}^{\mathrm{pressure}} + F_{i_f}^{\mathrm{viscous}} + F_{i_f}^{\mathrm{external}} + F_{i_f}^{\mathrm{g2f}})(t)$
        \EndFor
        \State Compute $\ddot{r}(t)$ and $\ddot{\theta}(t)$ with Eq.~\eqref{eq:tankdynamics}.
        \State \textbf{Return} $\dot x(t) = [\dot{r} \ \ \dot{\theta} \ \ \dot{r}_{i_f} \ \ \ddot{r} \ \ \ddot{\theta} \ \ \ddot{r}_{i_f}]^\top(t)$.
    \end{algorithmic}
\end{algorithm}

\subsection{Software implementation details\label{sec:software_implementation}}
We leverage the computational capabilities of the open-source software package \textsc{Jax} for \textsc{Python}~\cite{vanrossum2009} to implement Algorithm~\ref{alg:state_evol_fun} and the first-order symplectic Euler integrator, fulfilling the first objective defined in Subsection~\ref{sec:objectives}. The proposed implementation offers three main advantages:
\begin{enumerate}
    \item Algorithm~\ref{alg:state_evol_fun} requires iterative computations over all particles. The \textsc{Jax}-based implementation enables parallel execution of these operations on a GPU, such that the computational cost does not significantly increase with the number of fluid particles, up to the limits imposed by the available GPU memory.
    \item The use of \textsc{JAX} enables exact and efficient computation of the Jacobian of the state transition function via automatic differentiation.
    \item The implementation benefits from just-in-time (JIT) compilation, which translates high-level \textsc{Python} code into high-performance low-level instructions using the \emph{accelerated linear algebra} (XLA) compiler.
\end{enumerate}

To balance numerical accuracy and computational efficiency, a multi-rate simulation scheme is adopted. The fluid and rigid-body dynamics are integrated using a faster sampling rate to accurately capture the evolution of the fluid. In contrast, higher-level operations such as control input updates, data logging, Jacobian evaluations and visualization are performed at a lower sampling rate.

The computation of the correcting factor $\gamma_1$ in~\eqref{eq:gamma1} depends on several aspects, including the position of the fluid relative to the boundary and the local fluid density. In practice, a fixed value for $\gamma_1$ is used. This value can either be precomputed for a representative particle configuration reproducing the scenario shown in Fig.~\ref{fig:boundary_sketch}, or tuned empirically to prevent particle tunnelling during the simulation.

A wide range of kernel functions $W$ has been proposed in the SPH literature. In this work, we employ the cubic spline kernel~\cite{monaghan_2005}:
\begin{equation}\label{eq:cubicspline}
    W_{\mathrm{cb}}(x;h) = \frac{15}{14 \pi h^2}\begin{cases}(2-q)^3-4(1-q)^3, & \text { for } 0 \le q \le 1, \\ (2-q)^3, & \text { for } 1 \le q \le 2, \\ 0, & \text { otherwise},\end{cases} \qquad q = \frac{|x|}{h},
\end{equation}
for the computation of Eqs.~\eqref{eq:density_update},~\eqref{eq:momentum}, and~\eqref{eq:viscous}. Additionally, we use the spiky kernel introduced in~\cite{desbrun1996}, which has a non-vanishing gradient near the centre:
\begin{equation}\label{eq:spiky3}
    W_{\mathrm{s3}}(x;h) = \frac{10}{\pi h^5} \begin{cases}(h - x)^3, & \text { for } 0 \le x \le h, \\ 0, & \text {otherwise}, \end{cases}
\end{equation}
for the computation of Eqs.~\eqref{eq:pressure_b2f} and~\eqref{eq:viscous_b2f}. Note that both kernels~\eqref{eq:cubicspline} and~\eqref{eq:spiky3} are normalized for $x \in \Real^2$. 
\section{Learning of an LPV surrogate model~\label{sec:surrogate}}
In this section, we address the second objective defined in Subsection~\ref{sec:objectives}. Specifically, the identification of the model parameters $\vartheta$ that minimize the objective~\eqref{eq:pem}, together with the selection of a suitable model structure for $\hat{f}$ and $\hat{h}$.
To this end, we adopt an LPV-SS model structure with \emph{affine} dependence on the scheduling. This choice is motivated by three main considerations. First, the affine LPV structure retains a computational complexity comparable to that of linear models, making it well suited for rapid simulation. Second, this model structure offers utilization potential for control through the wide range of convex analysis and controller synthesis techniques. Third, the nonlinear scheduling map enables the extraction of complex nonlinear relationships from simulation data. While scheduling variables can be selected based on engineering insight, such relationships are difficult to derive for fuel sloshing due to the complex fluid--solid interactions. A data-driven scheduling approach is therefore adopted.

Then, the proposed parametrization of the surrogate model~\eqref{eq:surrogate} is given as
\begin{equation}\label{eq:surrogate_form}
    S_{\vartheta} : \left\{
    \begin{aligned}
        \hat{x}_{k+1} &= A(p_k) \hat{x}_k + B(p_k) u_k, \\
        \hat{y}_k &= C(p_k) \hat{x}_k + D(p_k) u_k,
    \end{aligned}
    \right.
\end{equation}
where $p_k \in \Real^{n_{\mathrm p}}$ is the scheduling variable that is defined through the \emph{scheduling map} $p_k = \eta(\hat{x}_k, u_k, \vartheta_\eta)$, where $\eta:\Real^{\hat{n}_{\mathrm{x}}} \times \Real^{\dnu} \times \Real^{n_{\vartheta_{\eta}}} \to \Real^{n_{\mathrm p}}$ is a parametrized  \emph{feedforward neural network} (FNN) representing a nonlinear function , as proposed in~\cite{bemporad2025}. This corresponds to a self-scheduled LPV model. Furthermore, the matrix functions $A,\ldots,D$, collected as
\begin{equation*}
    M(p_k, \vartheta_M) = \begin{bmatrix}
        A(p_k, \vartheta_M) & B(p_k, \vartheta_M) \\  C(p_k, \vartheta_M) & D(p_k, \vartheta_M)
    \end{bmatrix},
\end{equation*}
have an affine dependency on $p_k$:
\begin{equation}\label{eq:LPVparametrization}
    M(p_k, \vartheta_M) = M_0(\vartheta_M) + \sum_{i=1}^{n_{\mathrm p}} p_{k} M_i(\vartheta_M),
\end{equation}
which can also be interpreted as a linear output layer of the FNN $\eta$. Here, the parameters $\vartheta_M$ are the the elements of the matrices in \eqref{eq:LPVparametrization}.
Therefore, the parameter vector to learn becomes $\vartheta = \operatorname{vec}([M_0, M_1, \dots, M_{n_{\mathrm p}}], \vartheta_\eta) \in \Real^{n_\vartheta}$.

As the given dataset $\mathcal{D}_N$ is directly obtained from the fuel sloshing simulator, the measured data is noise-free. To address the identification problem with the prediction error minimization objective~\eqref{eq:pem}, we solve the following optimization proposed in~\cite{bemporad2025}:
\begin{equation}\label{eq:surrogate_optimization}
    \begin{aligned}
    \vartheta^\ast, \hat{x}_0^\ast &= \arg \min_{\vartheta, \, \hat{x}_0^\ast} J_{\D_N}(\vartheta, \hat{x}_0) + R(\vartheta, \hat{x}_0) \quad \text{subject to} \\
    &\hat{x}_{k+1} = A(p_k, \vartheta_M) \hat{x}_k + B(p_k, \vartheta_M) u_k, \\
    &\hat{y}_k = C(p_k, \vartheta_M) \hat{x}_k + D(p_k, \vartheta_M) u_k, \\
    &p_k = \eta(\hat{x}_k, u_k, \vartheta_\eta),\ k\in\{0,\ldots,N\},
    \end{aligned}
\end{equation}
where the loss $J_{\mathcal{D}_N}$ is defined in~\eqref{eq:pem}, $\hat{x}_0$ is the initial state and $R(\vartheta, \hat{x}_0)$ is a regularization term given by
\begin{equation}\label{eq:regularization}
    R(\vartheta, \hat{x}_0) = \frac{\sigma_2}{2}\|\vartheta\|_2^2 + \frac{\sigma_x}{2}\|\hat{x}_0\|_2^2,
\end{equation}
where $\sigma_2 > 0$ and $\sigma_x > 0$ are the weights corresponding to $\ell_2$ regularization of the model parameters and the initial state $\hat{x}_0$, respectively. The optimization~\eqref{eq:surrogate_optimization} is solved with a fixed number of Adam~\cite{kingma2017} gradient-descend steps that is used to warm-start a limited-memory Broyden-Fletcher-Goldfarb-Shanno (L-BFGS) scheme~\cite{byrd1995}, and the gradients are computed using automatic differentiation.
\section{Results~\label{sec:results}}
This section demonstrates the capabilities of the proposed SPH-based fuel sloshing simulator and evaluates the performance of an identified LPV surrogate model in a closed-loop attitude control setting under zero-gravity conditions. First, a benchmark scenario\footnote{The parameter values used for the fuel sloshing system are chosen to be physically plausible but not representative of a specific spacecraft, and atmospheric disturbances are neglected.} and attitude control setup are introduced. Second, closed-loop simulations under two distinct manoeuvre profiles are performed. Lastly, based on a separate open-loop excitation dataset, an LPV surrogate model is identified and validated on the closed-loop manoeuvre profiles. The software implementation of the simulator is publicly available at \href{https://gitlab.com/Javi-Olucha/fuel-sloshing-sph-simulator}{https://gitlab.com/Javi-Olucha/fuel-sloshing-sph-simulator}.
\subsection{Benchmark scenario and control setup}\label{sec:benchmark}
We consider a spacecraft consisting of a rectangular rigid body containing a circular fuel tank located at the CoM. The geometric and physical parameters of the satellite are summarized in Table~\ref{tab:satellite_params}. The tank is partially filled to 60\% with a fluid whose properties are selected to be representative of Hydrazine.
The fluid is discretized using 666 particles with a diameter of $6~\mathrm{mm}$, and a smoothing length of $9.42~\mathrm{mm}$ is used for the SPH kernel functions. The inner boundary of the tank is modelled using a single layer of 236 ghost particles, uniformly distributed along the tank wall. The corresponding fuel sloshing parameters are summarized in Table~\ref{tab:fuel_params}.
No gravity or atmospheric disturbance forces are considered.
The system inputs and outputs are defined as in Section~\ref{sec:scenario}, where the thruster forces and torque act at the CoM and the measured outputs correspond to the translational and rotational motion of the satellite CoM.

Both the satellite and the fluid are initialized at rest. To obtain a steady-state fluid configuration, a preliminary simulation is performed in which the fuel particles are spawned randomly distributed within the tank and allowed to evolve without external actuation until their velocities converge to zero. 
Lastly, for numerical simulation, a time step of $T_{\mathrm{s}} = 0.001~\mathrm{s}$ is used for integrating the system dynamics, while a time step of $T_{\mathrm{s}} = 0.05~\mathrm{s}$ is used for measuring, update the control action and update the visualization.
\begin{table}
\centering
\caption{Physical parameters of the considered satellite model.}\label{tab:satellite_params}
\small
\renewcommand{\arraystretch}{1.1}
\setlength{\tabcolsep}{5pt}
\begin{tabular}{l l l l}
\toprule
\textbf{Parameter} & \textbf{Value} & \textbf{Parameter} & \textbf{Value} \\
\midrule
Satellite geometry & Rectangular & Fuel tank geometry & Circular \\
Satellite width ($\mathrm m$) & $1.0$ & Satellite CoM ($\mathrm m$) & $(0, \, 0)$ \\
Satellite height ($\mathrm m$) & $0.5$ & Fuel tank CoM ($\mathrm m$) & $(0, \, 0)$\\
Satellite mass ($\mathrm{ kg}$) & 
$1010.71$ & Fuel tank inner wall radius ($\mathrm m$) & $0.2$ \\
Satellite inertia ($\mathrm {kg\cdot m^2}$) & 
$133.84$ & Fuel tank fill ratio (\%) & $60$ \\
\bottomrule
\end{tabular}
\end{table}
\begin{table}[t]
\centering
\caption{Physical parameters to model the considered fuel sloshing fluid.}\label{tab:fuel_params}
\small
\renewcommand{\arraystretch}{1.1}
\setlength{\tabcolsep}{5pt}
\begin{tabular}{l l l l}
\toprule
\textbf{Parameter} & \textbf{Value} & \textbf{Parameter} & \textbf{Value} \\
\midrule
Number of fluid particles & $666$ & Number of ghost particles & $236$ \\
Fuel particle length ($\mathrm{mm}$) & $6$ & Ghost particle length ($\mathrm {mm}$) & $5.4$ \\
Kernel smoothing length $h$ ($\mathrm{mm}$) & $9.42$ & Stiffness coefficient $k$ (-) & $3.0$\\
Base density $\rho_0$ ($\mathrm{kg/m^2}$) & $1017$ & Viscous factor $\alpha$ (-) & $8.32 \times 10^{-4}$ \\
Boundary viscous factor $\beta$ (-) & $4 \times 10^{-4}$ & Correcting factor $\gamma_1$ (-) & 0.5 \\
\bottomrule
\end{tabular}
\end{table}
\subsubsection{Attitude controller}
The angular motion of the spacecraft is regulated by a feedback attitude controller, while the translational motion remains open-loop. A negative feedback controller $K$ is designed following the guidelines in~\cite{guy2014}, with parameters tuned based on the static inertia of the satellite and the desired closed-loop response.
Specifically, the control law is given by
\begin{equation}
    \tau_k = K \ \begin{bmatrix}
        \tilde{\theta}_k - \theta_k & -\dot{\theta}_k
    \end{bmatrix}^\top, \quad K = \begin{bmatrix}
        J\omega^2 & 2 \xi J \omega
    \end{bmatrix},
\end{equation}
where $\theta_k$ and $\dot{\theta}_k$ are the sampled angular position and velocity, respectively, $\tilde{\theta}_k$ is the sampled angular reference signal, and $J$ is the static inertia of the satellite. The controller parameters are selected to achieve a closed-loop bandwidth of $0.1~\mathrm{Hz}$ and a damping ratio of $0.7$, corresponding to $\omega = 0.2 \pi$ and $\xi = 0.7$. Lastly, the controller output is kept constant during the sampling period by a zero order hold.

\subsubsection{Manoeuvrer profiles}
Two distinct manoeuvre profiles are defined to evaluate the simulator performance and to excite different fuel sloshing dynamics.
The first profile consists of a constant thrust along the $\mathrm{x}$-axis, representing a translational acceleration of the spacecraft. The performance of the attitude controller is then evaluated through a reference change in the desired angular position. Next, a pulsed force is applied along the $\mathrm{y}$-axis, representing an external disturbance (e.g., a micro-meteoroid impact), which induces a pendulum-like sloshing motion of the fluid.
The second profile consists of pulsed thrust inputs applied along both the $\mathrm{x}$ and $\mathrm{y}$-axes, followed by thrust in the opposite direction after a dwell time. This manoeuvre represents a diagonal translation typical of rendezvous operations, and excites a mass--spring--damper-like sloshing response.
\begin{figure}[t]
    \centering
    \includegraphics[width=1\linewidth]{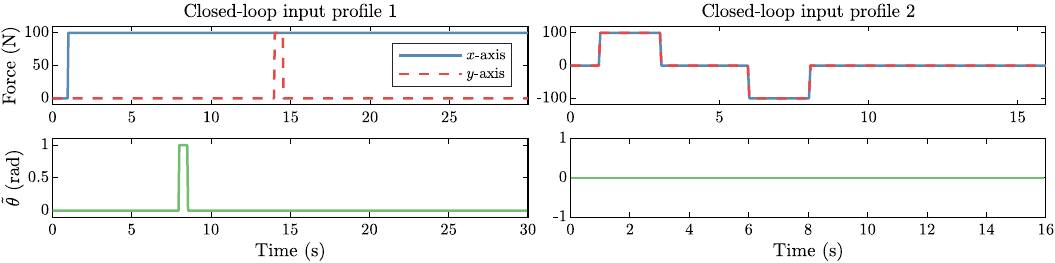}
    \caption{Manoeuvre profiles used in the closed-loop simulations. The desired angular position $\tilde{\theta}$ of the spacecraft is regulated by the controller $K$.}
    \label{fig:cl_profiles}
\end{figure}
Both input profiles are illustrated in Fig.~\ref{fig:cl_profiles}. 

\subsection{Closed-loop simulation of the spacecraft under fuel sloshing}\label{sec:cl_sloshing}
The benchmark scenario described in the previous Subsection~\ref{sec:benchmark} is simulated in closed loop for the two manoeuvre profiles to analyse the performance of the proposed SPH-based fluid simulator and investigate the coupled rigid-body and fluid dynamics under representative operating conditions.

\begin{figure}[t]
    \centering
    \includegraphics[width=1\linewidth]{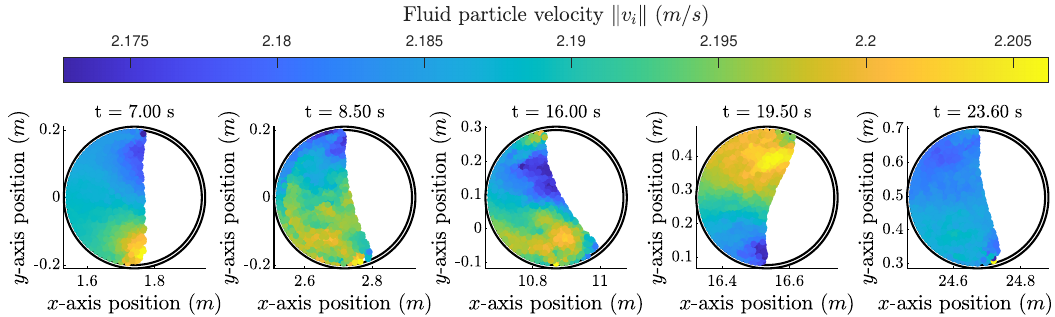}
    \caption{Visualization of the fluid particle distribution at different time snapshots for the first manoeuvre profile. A pendulum-like sloshing mode is observed following the disturbance along the $\mathrm{y}$-axis. The visualization of the spacecraft rigid body is omitted for clarity. A video of the corresponding simulation is available at \href{https://youtu.be/n0erMbdLwgc}{https://youtu.be/n0erMbdLwgc}.}
    \label{fig:pendulum-slosh}
\end{figure}

\begin{figure}[t]
    \centering
    \includegraphics[width=1\linewidth]{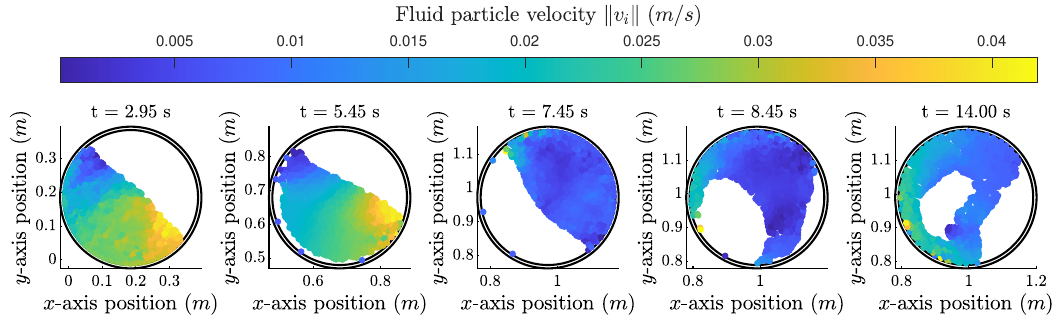}
    \caption{Visualization of the fluid particle distribution at different time snapshots of the closed-loop simulation for the second manoeuvre profile. The fluid exhibits a mass--spring--damper-like response, interacting with the tank walls during changes in the spacecraft acceleration. The visualization of the spacecraft rigid body is omitted for clarity. A video of the corresponding simulation is available at \href{https://youtu.be/SMiOcJaOFTk}{https://youtu.be/SMiOcJaOFTk}.}
    \label{fig:pendulum-msd}
\end{figure}

The simulation computation time\footnote{On a laptop with an i7-13850HX (2.10 GHz) CPU, an RTX2000 Ada (8 GB VRAM) GPU, and 64 GB RAM.} is reported in Table.~\ref{tab:cl_results}, and the simulated closed-loop trajectories of the spacecraft are shown in Fig.~\ref{fig:cl_results}.
It can be observed that for both manoeuvre profiles the attitude controller stabilizes the spacecraft effectively. 
Specifically, the first manoeuvre profile induces a pendulum-like behaviour on the fluid, which is shown via snapshots of the fluid particle distribution in Fig.~\ref{fig:pendulum-slosh}.
In contrast, as shown in Fig.~\ref{fig:pendulum-msd}, the second manoeuvre profile induces a mass--spring--damper-like behaviour on the fluid, interacting with the tank walls during changes in the spacecraft acceleration.

\begin{figure}[t]
    \centering
    \includegraphics[width=1\linewidth]{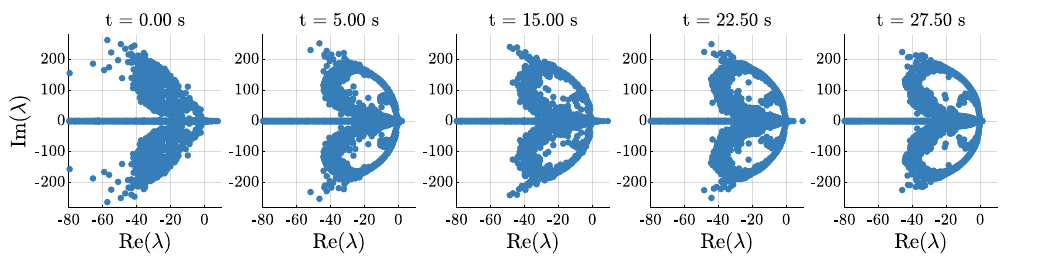}
    \caption{Eigenvalue evolution of the linearized system at different time instances of the closed-loop simulation for the first manoeuvre profile.}
    \label{fig:eig_profile1}
\end{figure}

\begin{figure}[t]
    \centering
    \includegraphics[width=1\linewidth]{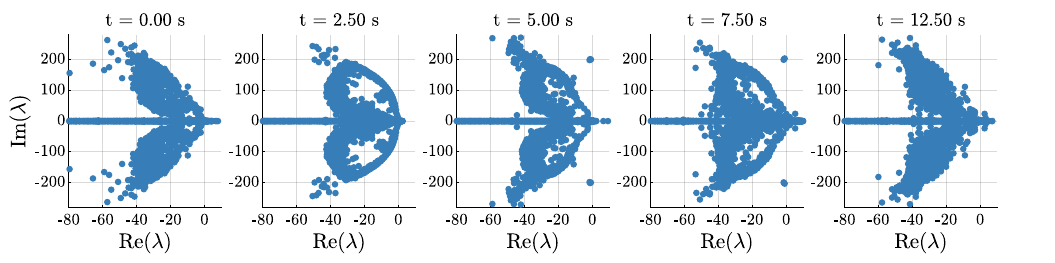}
    \caption{Eigenvalue evolution of the linearized system at different time instances of the closed-loop simulation for the second manoeuvre profile.}
    \label{fig:eig_profile2}
\end{figure}

To further analyse the nonlinear nature of the system dynamics, linearizations of the continuous-time nonlinear dynamic model are computed along the simulated trajectories using automatic differentiation. These linearizations are performed on the open-loop system dynamics, i.e., excluding the controller. 
The eigenvalues of the resulting linearized systems at different time instances are shown in Fig.~\ref{fig:eig_profile1} and Fig.~\ref{fig:eig_profile2} for the first and second manoeuvre profiles, respectively. The observed evolution of the eigenvalues over time highlights the inherently nonlinear nature of the system dynamics, which naturally motivates the use of a linear parameter-varying (LPV) framework.
\subsection{Learning an LPV surrogate model}
\subsubsection{Experiment design}
\begin{figure}[b]
    \centering
    \includegraphics[width=1\linewidth]{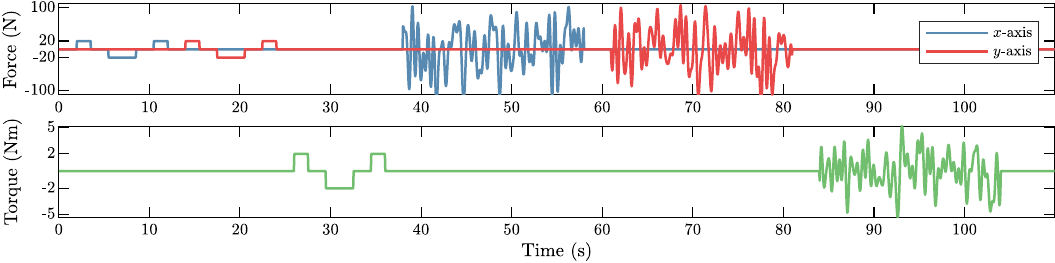}
    \caption{Input train signal used for system identification. The signal combines pulsed inputs and multi-sine components with random phase, applied independently to each input channel.}
    \label{fig:iden_profile}
\end{figure}
To address the LPV system identification problem, a training dataset distinct from the previously defined manoeuvre profiles is generated.
An input train signal, shown in Fig.~\ref{fig:iden_profile}, is designed with a length of 2200 samples and consists of a combination of pulsed inputs and multi-sine signals with random phase, applied independently to each input channel. The pulsed inputs are introduced to excite transient dynamics associated with mass--spring--damper-like sloshing behaviour. In addition, the multi-sine signals cover a frequency range of $[0,\ 2)~\mathrm{Hz}$ with a resolution of $0.05~\mathrm{Hz}$, exciting a wide spectrum of the system dynamics.
The designed input train signal is applied to the system in open-loop, and the resulting input--output trajectories are collected as the training dataset $\mathcal{D}_N$.

\subsubsection{Identification of an LPV surrogate model}
Based on the training dataset $\mathcal{D}_N$, our second objective is to identify an LPV surrogate model that provides a computationally efficient approximation of the spacecraft dynamics. The surrogate model should capture the input--output behaviour of the system, relating the applied forces and torque to the translational and rotational motion of the spacecraft.
A direct identification of the full input--output model, including positions and velocities, was found to be challenging. In particular, standard LTI identification approaches were not able to reliably capture the integrator dynamics associated with the position states. To address this issue, the identification problem is reformulated at the velocity level.
Specifically, a model that maps the input forces $u(t) =[u_{\mathrm x}(t) \ u_{\mathrm y}(t) \ \tau(t)]^\top$ to the output velocities $y(t) = [\dot{r}_{\mathrm x}(t) \ \dot{r}_{\mathrm y}(t) \ \dot{\theta}(t)]^\top$ is identified.
The full surrogate model, including position states, is subsequently recovered by augmenting the identified model with integrators.

For this purpose, we proceed to identify a DT LPV--SS model with affine dependence on the scheduling variable, as described in Section~\ref{sec:surrogate}. The model structure takes the form~\eqref{eq:surrogate_form} with sampling time $T_{\mathrm{s}} = 0.05~\mathrm{s}$, and dimensions $\hat{n}_{\mathrm x} = 4$, $n_{\mathrm u} = 3$, $n_{\mathrm y} = 3$, and $n_{\mathrm p} = 1$ for the state, input, output, and scheduling variables, respectively. 
The feedthrough matrix is constrained to zero, and the scheduling map $\eta(\hat{x}_k, u_k, \vartheta_\eta)$ is parametrized by a FNN with two hidden layers of four neurons each and $\mathrm{tanh}$ activation functions. The resulting LPV model contains a total of $n_\vartheta = 130$ trainable parameters.

The LTI component of the model, corresponding to $M_0$, is initialized using an DT LTI--SS model identified with the \textsc{ssest} function in \textsc{Matlab}. For this, the feedthrough matrix is fixed to zero, the estimation of a noise model is disabled, and the identification focus is set to simulation. The identified LTI model achieved an average \emph{best fit rate\footnote{\small $\mathrm{BFR} = \left(1 - \sqrt{\frac{\sum_{k=0}^{N} \|y(k) - \hat{y}(k) \|_2^2}{\sum_{k=0}^{N} \|y(k) - y_{\mathrm{mean}} \|_2^2}}\right)\cdot 100\%$, where $y$ is the data sequence, $y_{\mathrm{mean}}$ is the sample mean of $y$, and $\hat{y}$ is the predicted response of the model.}} (BFR) of $82.16\%$ on the training dataset. 

Next, the scheduling-dependent matrices $M_i$ are initialized from a zero-mean normal distribution, and the neural network weights are initialized using the Xavier method~\cite{glorot2010}. The regularization weights are set to $\sigma_2 = 10^{-4}$ and $\sigma_{\mathrm x} = 10^{-6}$.
The LPV model parameters are estimated by minimizing~\eqref{eq:surrogate_optimization} with the scaled and normalized dataset, using 2000 iterations of the ADAM optimizer followed by up to 6000 iterations of L-BFGS. The optimization is repeated eight times from different random initial guesses, resulting in a total training time of approximately $48~\mathrm{s}$. The final model is selected as the one achieving the highest average BFR on the training dataset, yielding an average BFR of $98.66\%$.
Finally, the full LPV surrogate model is obtained by augmenting the identified velocity-level model, resulting in
\begin{equation}
    S: \left\{ \begin{aligned}
        &\begin{bmatrix}
            \hat{x}^{\mathrm{e}}_{k+1} \\
            \hat{x}_{k+1}    
        \end{bmatrix}  = \begin{bmatrix}
            I & T_{\mathrm s} C(p_k) \\
            0 & A(p_k)
        \end{bmatrix}
        \begin{bmatrix}
            \hat{x}^{\mathrm{e}}_k \\
            \hat{x}_k
        \end{bmatrix} +
        \begin{bmatrix}
            T_{\mathrm s} D(p_k) \\
            B(p_k)
        \end{bmatrix} u_k, \\
        &\begin{bmatrix}
            \hat{y}_k^{\mathrm e} \\
            \hat{y}_k
        \end{bmatrix} = \begin{bmatrix}
            I & 0 \\
            0 & C(p_k)
        \end{bmatrix}\begin{bmatrix}
            \hat{x}^{\mathrm{e}}_k \\
            \hat{x}_k
        \end{bmatrix} + 
        \begin{bmatrix}
            0 \\
            D(p_k)
        \end{bmatrix} u_k,
    \end{aligned} \right.
\end{equation}
where $I$ and $0$ are identity and zero matrices of appropriate dimensions, and the augmented states $\hat{x}^{\mathrm{e}}_k \in \mathbb{R}^3$ and outputs $\hat{y}^{\mathrm{e}}_k \in \mathbb{R}^3$ correspond to the reconstructed position of the spacecraft.
%
% \begin{equation}
%     \begin{pmatrix}
%         x_{k+1} \\
%         z_k \\
%         y_k
%     \end{pmatrix} = 
%     \begin{pmatrix}
%         A & B_1 & B_2 \\
%         C_1 & 0 & D_{12} \\
%         C_2 & D_{21} & D_{22}
%     \end{pmatrix}
%     \begin{pmatrix}
%         x_k \\ w_k \\ u_k
%     \end{pmatrix}, \quad w_k = \Delta(p_k) z_k.
% \end{equation}
% \begin{equation*}
%     y_k = [v_k^x \ v_k^y \ v_k^\theta]^\top, \quad [q_{k+1}^x \ q_{k+1}^y \ q_{k+1}^\theta]^\top = [q_{k}^x \ q_{k}^y \ q_{k}^\theta]^\top + T_{\mathrm s}[v_k^x \ v_k^y \ v_k^\theta]^\top, 
% \end{equation*}
% \begin{equation*}
%     x_{k+1}^{\mathrm{e}} = y_k.
% \end{equation*}
% \begin{equation}
%     \begin{pmatrix}
%         x_{k+1}^{\mathrm{e}} \\
%         x_{k+1} \\ \hline
%         z_k \\ \hline
%         y_{k}^{\mathrm e} \\
%         y_k
%     \end{pmatrix} = 
%     \left(\begin{array}{cc|c|c}
%         I & T_{\mathrm s} C_2 & T_{\mathrm s} D_{21} & T_{\mathrm s} D_{22}\\
%         0 & A & B_1 & B_2 \\ \hline
%         0 & C_1 & 0 & D_{12} \\ \hline
%         I & 0 & 0 & 0 \\
%         0 & C_2 & D_{21} & D_{22}
%     \end{array}\right)
%     \begin{pmatrix}
%         x_{k}^{\mathrm{e}} \\ x_k \\ \hline w_k \\ \hline u_k
%     \end{pmatrix}, \quad w_k = \Delta(p_k) z_k.
% \end{equation}
\begin{table}[t]
\centering
\caption{Closed-loop simulation results. The reported computation time corresponds exclusively to the simulation of the system dynamics, excluding data storage, gradient computation and visualization.}\label{tab:cl_results}
\small
\begin{tabular}{l c c c c c c c c}
\toprule
\multirow{2}{*}{\textbf{Model}} & \textbf{Manoeuvre} & \textbf{Computation } &\multicolumn{6}{c}{\textbf{BFR} ($\%$)}  \\
& \textbf{profile} & \textbf{time} ($\mathrm s$)& $r_{\mathrm x}$ & $r_{\mathrm y}$ & $\theta$ & $\dot{r}_{\mathrm x}$ & $\dot{r}_{\mathrm y}$ & $\dot{\theta}$ \\ \midrule
\multirow{2}{*}{SPH simulator}  & 1 & 9.9093 &  &  &  &  &  &   \\
 & 2 & 5.1876 &  &  &  &  &  &   \\
\multirow{2}{*}{Identified LTI} & 1 & 0.0038 & 84.41 & -146.95 & 96.40 & 71.85 &  -99.11 & 83.23  \\
 & 2 & 0.0023 & 45.92 & 93.02 & -436.92 & 72.66 & 88.38 & -557.13 \\
\multirow{2}{*}{Identified LPV} & 1 & 0.0242 & 99.52 & 95.60 & 98.18 & 98.94 & 94.64 & 90.99  \\
 & 2 & 0.0056 & 99.40 & 98.62 & -66.01 & 97.17 & 96.61 & -41.68 \\
\bottomrule
\end{tabular}
\end{table}
\subsection{Closed-loop validation of the identified surrogate models}
The identified LTI and LPV models are validated in closed-loop simulations using the two manoeuvre profiles introduced in Section~\ref{sec:benchmark}. Table~\ref{tab:cl_results} summarizes the results in terms of average best fit ratio (BFR) and computation time. The reported computation time corresponds exclusively to the simulation of the system dynamics.

\begin{figure}[t]
    \centering
    \includegraphics[width=1\linewidth]{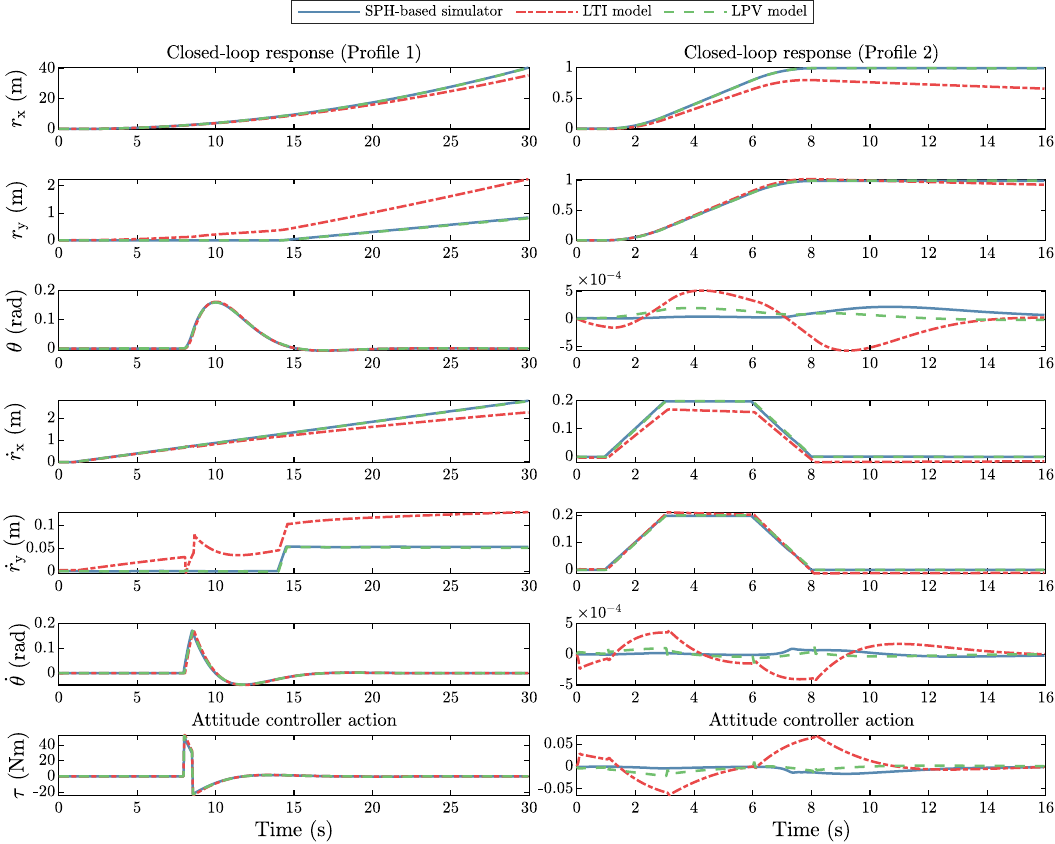}
    \caption{Closed-loop simulation results between the SPH-based simulator and the identified LTI and LPV models for the two manoeuvre profiles.}
    \label{fig:cl_results}
\end{figure}

The simulation results, shown in Fig.~\ref{fig:cl_results}, indicate that the LPV surrogate model consistently reproduces the dominant dynamics of the spacecraft with high accuracy across both manoeuvre profiles, with BFR values close to $100\%$ for most outputs. The lower BFR observed for the angular motion in the second profile is attributed to the small angular rate, on the order of $10^{-4}$, which results in a negligible contribution to the overall system dynamics. In contrast, the performance of the LTI surrogate model is significantly lower, particularly for manoeuvre profile~2, highlighting the limitations of LTI models in capturing the inherently nonlinear nature of the fuel sloshing dynamics.
In terms of computational efficiency, both surrogate models provide a substantial reduction in simulation time compared to the SPH-based simulator. In particular, the LPV surrogate achieves simulation times that are more than two orders of magnitude faster, while maintaining high accuracy. Although the LTI model is faster, this comes at the cost of reduced accuracy.

These results demonstrate that the LPV surrogate model provides a reliable and computationally efficient approximation of the dominant spacecraft dynamics, which makes the proposed LPV surrogate model particularly suitable for applications requiring repeated simulations.
In particular, the LPV surrogate can be used to accelerate Monte Carlo simulation campaigns and to rapidly screen or discard unsuitable controller designs before validation with the high-fidelity model.
\section{Conclusions~\label{sec:conclusions}}
In this paper, we have presented an open-source, high-fidelity SPH-based fuel sloshing simulator capable of rapidly generating informative datasets, including system inputs, outputs, states, and linearized dynamics, to support surrogate model extraction. Leveraging the computational capabilities of \textsc{Jax}, the proposed implementation exploits GPU parallelization, and the computational cost of the simulator scales favourably with the number of fluid particles. This enables efficient analysis and data generation across a wide range of simulation scenarios.
As demonstrated in Subsection~\ref{sec:cl_sloshing}, the simulator reproduces the expected fuel sloshing behaviour of a partially filled tank in a spacecraft under closed-loop attitude control, achieving a computation time below $10~\mathrm{s}$ for a 30~s simulation for a benchmark consisting of 666 fluid particles.
Furthermore, we have demonstrated that LPV surrogate models can be used to effectively capture the fuel sloshing dynamics. The identified LPV surrogate model has been validated in closed loop, where it accurately reproduces both pendulum--like and mass--spring--damper-like sloshing modes, while achieving simulation times that are more than $100\times$ faster than the high-fidelity model.
For further research, we plan to investigate on leveraging linearized dynamic information during the LPV surrogate extraction process, and on characterizing the surrogate model uncertainty.
% % % % % % % % % % % % % % % % % % % % % % % % % % % % % % % % % % % % % % % % 
\section*{Declaration of Use of Artificial Intelligence}
We have used learning based methods to estimate models from data with a \textsc{Python} implementation. No other forms of artificial intelligence have been used in the work presented.
\bibliography{EuroGNC2026}

\end{document}